\documentclass{elsart3p}
\usepackage{graphicx,amsmath,amssymb,mathptmx}

\journal{Physics Letters A}

\newcommand{\dr}{\mathrm{d}}
\DeclareMathOperator{\sech}{sech}
\DeclareMathOperator{\csch}{csch}

\begin{document}
\begin{frontmatter}

\title{Physical dynamics of quasi-particles in nonlinear wave equations}

\author{Ivan Christov\thanksref{tamu}\corauthref{cor}\thanksref{nu}}, 
\ead{\href{mailto:christov@alum.mit.edu}{christov@alum.mit.edu}}
\author{C. I. Christov\thanksref{ull}}
\ead{\href{mailto:christov@louisiana.edu}{christov@louisiana.edu}}
\ead[url]{\url{http://www.ucs.louisiana.edu/~cic6380/}}
\address[tamu]{Department of Mathematics, Texas A\&M University, College Station, TX 77843-3368, USA}
\thanks[nu]{Present address: Department of Engineering Sciences and Applied Mathematics, Northwestern University, Evanston, IL 60208-3125, USA.}
\corauth[cor]{Corresponding author.}
\address[ull]{Department of Mathematics, University of Louisiana at Lafayette, Lafayette, LA 70504-1010, USA}

\begin{abstract}
By treating the centers of solitons as point particles and studying their discrete dynamics, we demonstrate a new approach to the quantization of the soliton solutions of the sine-Gordon equation, one of the first model nonlinear field equations. In particular, we show that a linear superposition of the non-interacting shapes of two solitons offers a qualitative (and to a good approximation quantitative) description of the true two-soliton solution, provided that the trajectories of the centers of the superimposed solitons are considered \emph{unknown}. Via variational calculus, we establish that the dynamics of the quasi-particles obey a pseudo-Newtonian law, which includes cross-mass terms. The successful identification of the governing equations of the (discrete) quasi-particles from the (continuous) field equation shows that the proposed approach provides a basis for the passage from the continuous to a discrete description of the field.
\end{abstract}

\end{frontmatter}

\section{Introduction}
\label{sec:intro}
A sought-after property of model field equations is that they possess localized, permanent wave solutions that retain their identity upon interactions with one another. \emph{Solitons}, which are solutions of \emph{fully-integrable} equations, are an example of such waves. Unfortunately, integrability is not always a property of models that are of physical importance. Therefore, it is important to develop simple, albeit approximate, approaches to studying the dynamics of permanent waves in non-integrable systems, for which exact solutions are difficult to obtain.

The idea of identifying a localized solution of a nonlinear wave equation with an elementary particle was first proposed by Perring \& Skyrme \cite{PerringSkyrme}. They found a solution of (what is known today as) the sine-Gordon equation (SGE) \cite{Rubinstein} consisting of two interacting, localized waves, which is an example of (what is nowadays called) a \emph{two-soliton solution}, several years before the notion of a soliton was introduced by Zabusky \& Kruskal \cite{Zabu65}. Through numerical simulation, the latter authors discovered that the localized traveling-wave solutions of the Korteweg--de Vries  equation retain their shapes (identities) after they pass through each other (interact). Apparently, Zabusky \& Kruskal were unaware of the work of Perring \& Skyrme \cite{PerringSkyrme} and arrived at the idea of identifying the nonlinear waves' dynamics with those of particles through experiment rather than conjecture. 

Since their discovery, solitons have attracted an enormous amount of attention. Significant progress has been made in their mathematical description, and their applications have been far-reaching \cite{Filippov,Dauxois06}. The so-called kink solitary waves considered by Perring \& Skyrme \cite{PerringSkyrme} have been shown to be indeed solitons \cite{Dauxois06,BulloughCaudrey}. Moreover, the relationship between the particle-like dynamics of the coherent structures that emerge in the solutions of nonlinear wave equations and the field theories of particle physics are well-established in the literature \cite{MauginChristov,Filippov,KaupNewell,BowtellStuart}. 

From a mathematical point of view, we can elucidate the latter relationship by, somehow, reducing the ``infinitely complex'' continuous description of the field to a ``finitely complex'' discrete description. In this Letter, we show how this can be achieved by studying the dynamics of \emph{quasi-particles}. Our approach amounts to ``degrading'' the continuous description of the wave profile to a discrete description of the centers of coherent structures, assuming that the shapes of the coherent structures (for an integrable system, this would simply be the solitons) are not significantly affected by the each other's presence. To this end, in this Letter, the term ``quasi-particle'' refers to these permanent, indestructible and virtually non-deformable coherent structures, whose centers can be treated as point particles with mass equal to some associated measure of inertia that we call the \emph{pseudomass}. Furthermore, replacing a complicated continuous profile by a superposition of localized shapes can also be viewed as a \emph{coarse-grain description}, in the sense that small deformations and wiggles are filtered out from the profile leaving merely the main structure (the ``grains'').

Hence, the coarse-grain description amounts to replacing the solution of a nonlinear wave equation with a linear superposition of basis states (i.e., traveling-wave solutions in their undeformed, or non-interacting, state), then the trajectories of the centers of the superimposed waves, which for a nonlinear equation do not follow the undisturbed (linear) trajectories, are treated as unknown. Restricting to the case of just two superimposed waves, a discrete model for the  trajectories is derived and solved numerically in this Letter. This approach gives a wave profile whose deviation from the analytical two-soliton solution (when available) is orders of magnitude smaller than the characteristic size of the solitons. Consequently, this paves the way to constructing successive approximations that account for the higher-order, nonlinear interactions of solitons.

Here, we note that our coarse-grain description is a special interpretation of the more general method of \emph{collective coordinates/variables}, which has been put on solid theoretical ground \cite{Willis88} and become part of the textbooks on solitons \cite{Dauxois06}. The latter approach made its debut in the study of resonances and collisions of solitary waves in the so-called $\phi^4$ equation \cite{Sugiyama,Campbell83}, a close relative of the SGE, and the study of two-soliton interactions in the SGE \cite{Karpman,Willis98}. Moreover, the method of collective variables is just one type of \emph{variational approximation}, which is another approach to the analytical study of nonlinear wave equations that has recently regained popularity \cite{KaupMalomed_ZS,MalomedWeinstein,Kaup05}. Furthermore, it appears that Rice \cite{Rice} was the first to realize that the collective-coordinate variational approximation provides a way of ``distilling'' the particle-like dynamics of nonlinear waves from the continuos (field) description, though Karpman \& Solov'ev \cite{Karpman} had the foresight to use the term `quasi-particle' in their discussion.

Finally, we note that the nonlinear wave equation featured herein --- the sine-Gordon equation --- continues to be of interest as a model field equation \cite{LouHuTang05}. In addition, various modifications of it have been considered in the literature. For example, in order to establish the effects of acceleration on the shape of the SGE's solitons, Fogel et al.~\cite{Fogel77} introduced a driving force into the SGE. However, this required also adding dissipation in the field equation in order to stabilize the evolution of the solitons, i.e., to ensure that they reach a steady terminal velocity \cite{Fogel77,ReinFern}. Adding dissipation opens new horizons of investigation, and different physical mechanisms can be considered as progenitors of the dissipative force. It is well known that, in fluid mechanics, linear dissipation of either viscous or Darcy type can balance the nonlinearity in the field equation, and allow stable localized waves to exist \cite{Jordan04,Jordan06}. Moreover, in incompressible shallow-water flows, a viscous dissipation can allow for the existence of localized coherent structures with solitonic behavior \cite{ChriVel_PhysD}. Nonetheless, one thing is certain: dissipation can alter the behavior of a nonlinear field equation dramatically. Therefore, in this Letter, we focus on the \emph{lossless} SGE of Perring \& Skyrme \cite{PerringSkyrme} and show that the coarse-grain description of the field leads to the physical dynamics of \emph{locally-accelerating} quasi-particles, \emph{without}  introducing a driving force or dissipation into the equation. In this respect, however, the SGE differs fundamentally from the modern lossless nonlinear field theories of continuum mechanics (see, e.g., Ref.~\cite{js07} and those therein), since the ``unbalanced'' nonlinearity in the former allows for the creation of  localized coherent structures and does not lead to formation of singularities in  finite time.

\section{The sine-Gordon equation and its soliton solutions}
\label{sec:sG}
For the purposes of this Letter, the SGE takes the following dimensionless ($\hbar=c=m=1$) form:
\begin{equation}
u_{tt} - u_{xx} = - \sin u,\label{eq:sG}
\end{equation}
where the subscripts denote partial differentiation. We have selected the SGE  as our featuring example because there are known analytical expressions for its two-soliton solutions. This allows us to show that the coarse-grain description is both an effective approximation tool and a new method for nonlinear-wave quantization.

It is easy to show that the Lagrangian and the Hamiltonian of the SGE read
\begin{equation}
L,H = \int_{-\infty}^{+\infty} \tfrac{1}{2} u_t^2 \mp \left(\tfrac{1}{2} u_x^2 + 1 - \cos u\right) \dr x, \label{eq:HL_L}
\end{equation}
respectively \cite{Rubinstein,Dauxois06}, where the ``$-$'' sign in the right-hand side refers to $L$ and the ``$+$'' sign to $H$. In addition, the \emph{wave momentum} is defined \cite{MauginChristov} as
\begin{equation}
P = -\int_{-\infty}^{+\infty} u_x u_t \,\dr x.
\label{eq:wave_momentum}
\end{equation}
Then, the conservation of the energy and linear momentum require that $\dr H/\dr t = 0$ and $\dr P/\dr t = 0$.

Now, if we consider the moving frame $\xi = x-vt$, Eq.~\eqref{eq:sG} reduces to an ODE, which has the following solution:
\begin{equation}
u = \phi(\xi) = 4 \arctan\biggl[\exp\biggl(\frac{\xi}{\sqrt{1-v^2}}\biggr)\biggr], \quad 0 \le v < 1.
\label{eq:1kink}
\end{equation}
Notice that the latter is a \emph{one-soliton solution} because $\tfrac{1}{2\pi}[\lim_{x\to\infty}u(x,t)-\lim_{x\to-\infty}u(x,t)] = 1$ for all $t<\infty$ \cite{Rubinstein}. The kink, whose shape is termed a hydraulic jump in fluid mechanics, is a localized wave in the sense that its derivative 
\begin{equation}
u_x = \frac{2}{\sqrt{1-v^2}}\sech\biggl(\frac{\xi}{\sqrt{1-v^2}}\biggr) 
\label{eq:derivative_of_kink}
\end{equation}
is a function of finite energy (i.e., it is square-integrable).

Another well-known analytical solution of the SGE is the \emph{two-soliton solution} \cite{BulloughCaudrey}, which can take two distinct forms:
\begin{subequations}
\label{eq:4pi_kink}
\begin{align}
u^+ &=  4 \arctan\biggl\{\frac{\cosh[(\theta_1-\theta_2)/2]}{a_{12}
\sinh[(\theta_1+\theta_2)/2]}\biggr\},\label{eq:4pi_kink_p}\\[2mm]
u^- &=  4 \arctan\biggl\{\frac{\sinh[(\theta_1-\theta_2)/2]}{|a_{12}|
\cosh[(\theta_1+\theta_2)/2]}\biggr\},\label{eq:4pi_kink_m}
\end{align}
\end{subequations}
where, for $j\in\{1,2\}$,
\begin{subequations}
\label{eq:a12}
\begin{align}
\theta_j &= \gamma_j(x - v_j t - x_j),\quad \gamma_j^2 = (1-v_j^2)^{-1}, \\ 
a_j^2&= (1-v_j)/(1+v_j), \\
a_{12} &= (a_1-a_2)/(a_1+a_2).
\end{align}
\end{subequations}
It is important to note that, due to the multivalued nature of the inverse tangent function, one has to be careful about the definition of $u^+$. The inverse tangent in Eq.~\eqref{eq:4pi_kink_p} should \emph{not} be taken in the ``principal value'' sense because, then, it leads to a discontinuous definition of $u^+$. But, if the quadrant of the argument of the inverse tangent is taken into account, then $u^+$ is well-defined. 

\begin{figure}
\centerline{\includegraphics[width=6.8cm]{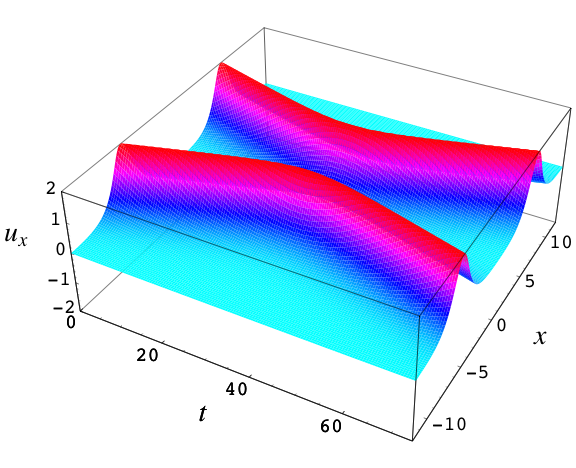}}
\centerline{\includegraphics[width=6.8cm]{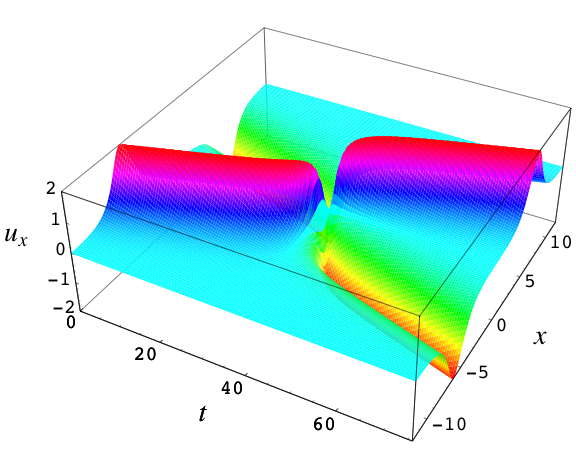}}
\caption{(Color online) Space-time plots of the soliton-soliton (top) and the soliton-antisoliton (bottom) interaction.}
\label{fig:sol-sol-anti}
\end{figure}

Henceforth, the solutions $u^+$ and $u^-$ are referred to as the \emph{soliton-soliton} (S-S) and \emph{soliton-antisoliton} (S-A) solutions, respectively. To justify this terminology, we present the spatial derivatives, which are the quantities of importance in defining the corresponding quasi-particles, of the S-S and S-A solutions in Fig.~\ref{fig:sol-sol-anti}.

Now, when the two solitons are far from each other (as $t\rightarrow+\infty$), it can be shown that Eq.~\eqref{eq:4pi_kink} reduces to
\begin{equation}
u^\pm \equiv 4 \arctan\bigl[\exp(\theta_1+\delta^\pm_1)\bigr] \pm
4 \arctan\bigl[\exp(\theta_2+\delta^\pm_2)\bigr], 
\label{eq:separated}
\end{equation}
where $\delta^\pm_1 = - \delta^\pm_2 = \frac{1}{2} \ln |a_{12}|$ are the \emph{phase shifts} \cite{BulloughCaudrey}. The above decomposition means that far from the point of collision of the two solitons, the profile is merely the \emph{linear} superposition of two single solitons [whose shapes are identical to the one-soliton profile given by Eq.~\eqref{eq:1kink}]. However, we cannot immediately express the two-soliton solution as a sum of two single solitons because of the phase shifts, which are, in fact, the most conspicuous manifestation of the nonlinearity as far as interacting solitons are concerned. Yet, the actual wave profile is approximated quite well by the mere sum of one-soliton profiles provided their centers follow trajectories that somehow account for the nonlinear interaction. The technical details and physical underpinning of this approximation are the goal of the this Letter.

Here, we must note that Bowtell \& Stuart \cite{BowtellStuart} developed a method, based on the poles of the Hamiltonian density, that allowed them to compute the exact (ODEs for the) trajectories of the centers of the solitons. Their technique has not been applied to any other well-known nonlinear wave  equations, which makes the SGE unique from the point of view of availability of reference solution for the trajectories of the centers to establish the practical applicability of the present approach. However, we emphasize again that we do not intend to develop an exact method for computing the trajectories based on particular properties of the governing nonlinear wave equation, rather we show how to approximate them based on more general principles. Thus, the method proposed below is applicable to integrable and non-integrable equations alike. 

\section{From continuous to discrete dynamics}
\subsection{One-soliton case}
To elucidate the concept of a quasi-particle we consider the propagation of a single solitary wave, whose shape is given by Eq.~\eqref{eq:1kink}, centered at certain spatial position $x=X(t)$ (termed the \emph{particle trajectory}), which  may be unknown in advance. Under the latter assumptions we have that
\begin{equation}
u(x,t) = \phi(\xi),\quad \xi = x - X(t),\quad v(t)\stackrel{\mathrm{def}}{=}\dot X(t).
\end{equation}
Consequently, the time derivative of wave profile is
\begin{equation}
u_t = 2\left[\frac{x-X(t)}{(\sqrt{1-v^2})^3}v\dot v - \frac{v}{\sqrt{1-v^2}}\right] \sech\biggl[\frac{x-X(t)}{\sqrt{1-v^2}}\biggr]. 
\label{eq:u_t}
\end{equation}

Now, the straightforward way to compute the discrete Lagrangian and Hamiltonian for the motion of the center of the soliton is to insert Eqs.~\eqref{eq:derivative_of_kink}~\&~\eqref{eq:u_t} into Eq.~\eqref{eq:HL_L} and to integrate over the infinite domain. However, in order not to obscure the main idea of the present work, we restrict ourselves to the non-relativistic (or  classical) approximation, i.e., $\dot X\equiv v\ll1$ $\Rightarrow v^2\lll 1$ (small phase speed) and $\ddot X \equiv\dot v = \mathcal{O}(v)$ (small phase acceleration). This means that the speed of the quasi-particle is much smaller than the characteristic speed of the equation (i.e., unity) and that the asymptotic order (of ``smallness'') of the acceleration is the same as that of the velocity. Under these assumptions the first term in Eq.~\eqref{eq:u_t} can be disregarded because it is of second order of ``smallness'' while the second term is of first order. In addition, any explicit dependence of the shape of the soliton on its phase speed is negligible in the classical limit. Finally, we note that the classical approximation is valid for speeds up to $v\simeq 0.2$. 

Thus, under the above assumptions, we get the following simple expressions for the profile and its partial derivatives:
\begin{equation}
\begin{aligned}
&u = 4 \arctan[\exp(\xi)],\quad 1-\cos u = 2 \sech^2\xi,\\
&u_x = 2 \sech\xi, \quad u_t = -2 \dot X \sech\xi.
\end{aligned}
\label{eq:cg_relations}
\end{equation}
Then, by substituting the latter expressions into Eq.~\eqref{eq:HL_L}, we get that the \emph{discrete} Lagrangian $\mathbb{L}$ is
\begin{equation}
\mathbb{L} = \tfrac{1}{2} \mathbb{M} \dot X^2, \quad \mathbb{M} \stackrel{\mathrm{def}}{=} 4\int_{-\infty}^{+\infty}\sech^2\xi \,\dr \xi = 8,
\end{equation}
which is identical to the Lagrangian of a particle of mass $\mathbb{M}$ undergoing inertial motion. Similarly, we obtain, by substituting Eqs.~\eqref{eq:cg_relations} into Eq.~\eqref{eq:wave_momentum}, that the \emph{discrete} wave momentum $\mathbb{P}$ is just 
\begin{equation}
\mathbb{P} = -\int_{-\infty}^{+\infty} -4\dot X \sech^2\xi \,\dr \xi 
= \mathbb{M} \dot X,
\label{eq:wave_momentum_discrete}
\end{equation}
which (again) is identical to the momentum of a (point) particle of mass $\mathbb{M}$ undergoing inertial motion. To distinguish between the properties of the continuous and discrete descriptions, we call $\mathbb{M}$ the \emph{pseudomass} and $\mathbb{P}$ the \emph{pseudomomentum}. Hence, Eq.~\eqref{eq:wave_momentum_discrete} shows that the wave momentum of a single solitary wave is the momentum of its ``center,'' if the latter is considered as a point particle endowed with mass $\mathbb{M}=8$. Of course, this should be of no surprise if the solutions to the SGE are to model elementary particles, which obey the principle of wave-particle duality.

So far we have used the term ``center of a soliton'' without defining it. Unfortunately, there is no general agreement on the matter, and the center of a coherent structure (e.g., a soliton) has to be defined on a case-by-case basis. A definition is given by Bergman et al.~\cite{Bergman83} but, as Kaup \cite{Kaup} shows, depending upon the definition being used, one may obtain that the quasi-particle's dynamics are Newtonian \cite{Fogel77} or non-Newtonian \cite{ReinFern} --- clearly, this is quite a discrepancy. In this respect, the closest work to ours is that of Rice \cite{Rice}, who derives the discrete Hamiltonian and Lagrangian for a single quasi-particle whose support can be considered an internal parameter of the problem. He comes to the conclusion that the dynamics of a deforming quasi-particle are, generally, non-Newtonian, as do we. For the purposes of this Letter, it suffices to define the center of a coherent structure $u$ as the point where $u_x$ attains its maximum (recall Fig.~\ref{fig:sol-sol-anti}), and that is precisely where the corresponding quasi-particle should be located. Note that, for the SGE solitons, this also corresponds to the inflection point of $u(\cdot,t)$, which is always located at $x^*$ such that $u_{xx}(x^*,t) = 0$.

\subsection{Two-soliton case}
Continuing on to the two-soliton case, we first note that one can formally  represent any two-soliton (two-wave, if the equation is non-integrable) solution of the field equations as the combination of the undeformed (non-interacting) shapes of two solitary waves $\Phi_1$, $\Phi_2$ and an arbitrary disturbance term $\Phi_{12}$ due to the interaction of the latter two:
\begin{multline}
u(x,t) = \Phi_1[x-X_1(t)] + \Phi_2[x-X_2(t)] \\ + \Phi_{12}[x-X_1(t),x-X_2(t)].\label{eq:coarse_grain}
\end{multline}
In fact, one can think of $\Phi_{12}$ as the difference between the full two-soliton solution given by Eq.~\eqref{eq:4pi_kink} and the linear superposition of two single soliton given by Eq.~\eqref{eq:separated}, with properly selected trajectories that are unknown in advance.

The two first terms of Eq.~\eqref{eq:coarse_grain} give a linear superposition that is bound to be a poor quantitative description of interaction of two nonlinear waves when they overlap significantly. Indeed, it is well-known that, e.g., for the Korteweg--de Vries equation, the maximal amplitude of a two-soliton solution is much smaller than the sum of the maximal amplitudes of the individual solitons \cite{Dauxois06}. But, a linear superposition along two undisturbed trajectories $X_i(t)=X_i(0)+v_it$ would give a maximal amplitude equal to the sum of the two maximal amplitudes. 

The main idea of the present work is that if the trajectories $X_i(t)$ are selected in a specific manner, then the predominant part of the nonlinear effects, represented by $\Phi_{12}$ in Eq.~\eqref{eq:coarse_grain}, in the two-soliton solution can be recovered by the linear superposition alone, i.e., Eq.~\eqref{eq:coarse_grain} with $\Phi_{12}$ neglected. That is to say, for properly chosen trajectories, one has $\max{|\Phi_{12}|} < \varepsilon \max\{\max|\Phi_1|,\max|\Phi_2|\}$ for some small $\varepsilon(>0)$. It is not know in advance how small $\varepsilon$ is, but one can estimate it \textit{a posteriori} and vindicate the approximation. We do so by comparison to the exact solution of the problem, but one can also judge the quality of the approximation, without \emph{any} knowledge of the exact solution, by the novel method of Kaup \& Vogel \cite{KaupVogel}. The latter will be necessary when the governing equation is non-integrable and the former approach fails.

Now, we show how the trajectories of the quasi-particles [as quantified by the functions $X_1(t)$ and $X_2(t)$] can be calculated. First, recall that we are working in the classical limit so we can neglect the dependence of a soliton's shape on its phase speed and use the undeformed (non-interacting) shape of the soliton in the calculation. Thus, after setting $\Phi_1=\Phi_2\stackrel{\mathrm{def}}{=}\Phi$, neglecting the term $\Phi_{12}$ in Eq.~\eqref{eq:coarse_grain} and acknowledging Eq.~\eqref{eq:separated}, we arrive at the following simple ensemble of two waves as the coarse-grain approximation: 
\begin{equation}
\begin{aligned}
u^\pm(x,t) &= \Phi[x-X_1(t)] \pm \Phi[x-X_2(t)], \\
\Phi(\zeta) &= 4\arctan[\exp(\zeta)],
\end{aligned}
\label{eq:coarse_grain_approx}
\end{equation}
where the ``$+$'' and ``$-$'' sign follow the convention set forth in Eq.~\eqref{eq:4pi_kink}. Then, the first partial derivatives of the wave profile $u^\pm$ are just
\begin{subequations}
\begin{align}
u_t^\pm(x,t) &= -\dot X_1 \Phi'[x-X_1(t)]  \mp \dot X_2 \Phi'[x-X_2(t)], \\
u_x^\pm(x,t) &= \phantom{-\dot X_1}\Phi'[x-X_1(t)] \pm \phantom{\dot X_2} \Phi'[x-X_2(t)].
\end{align}
\end{subequations}
Henceforth, we leave the time dependence of $X_1$ and $X_2$ implicit to simplify the notation. Consequently, the kinetic energy term of the Lagrangian, for both $u^+$ and $u^-$ (denoted simply by $u$), reads
\begin{multline}
\int_{-\infty}^{+\infty} \tfrac{1}{2}u_t^2 \,\dr x = \tfrac{1}{2} \mathbb{M}_{11} X_1^2 + \tfrac{1}{2} \mathbb{M}_{22} X_2^2 \\ + \mathbb{M}_{12}(X_2-X_1) \dot X_1 \dot X_2,
\end{multline}
where the pseudomasses $\mathbb{M}_{ij}$ ($i,j\in\{1,2\}$) are given by
\begin{subequations}
\begin{gather}
\begin{aligned}
\mathbb{M}_{12}(X_2-X_1) &\stackrel{\rm def}{=} \pm\int_{-\infty}^{+\infty}\Phi'(x-X_1)\Phi'(x-X_2) \,\dr x\\ 
&= \pm 4\int_{-\infty}^{+\infty} \sech(x -X_1)\sech(x-X_2) \,\dr x\\ 
&= \pm 8(X_2 - X_1)\csch(X_2 - X_1),
\end{aligned}\\
\mathbb{M}_{ii} \stackrel{\rm def}{=} 
\int_{-\infty}^{+\infty} [\Phi'(\xi_i)]^2 \,\dr x = 4 \int_{-\infty}^{+\infty} \sech^2\xi_i \,\dr\xi_i = 8,
\end{gather}
\end{subequations}
and $\xi_i = x - X_i$.

Similarly, we can compute the interaction potential (i.e., the remaining terms of the Lagrangian). First, we note that 
\begin{multline}
\tfrac{1}{2}u_x^2 + 1 - \cos u = 4\bigl\{\sech^2(x-X_1) + \sech^2(x-X_2) \pm \\ [\cosh(2x-X_1-X_2)-1] \sech^2(x-X_1)\sech^2(x-X_2)\bigr\}.
\label{eq:poteng}
\end{multline}
Then, by integrating the latter [Eq.~\eqref{eq:poteng}] over space\footnote{These integrals are rather difficult, thus we refer the reader to the appendices of Ferguson \& Willis  \cite{Willis98} and Sugiyama  \cite{Sugiyama} for explicit formul\ae\ and hints.}, we find that the potential energy of interaction is
\begin{equation}
U(z) \stackrel{\mathrm{def}}{=} 16 \pm 4\csch^3\bigl(\tfrac{1}{2}z\bigr)\sech\bigl(\tfrac{1}{2}z\bigr)(\sinh z - z),
\end{equation}
and we have set $z\stackrel{\mathrm{def}}{=}X_2-X_1$ for convenience. Consequently, combining the above computations, we obtain the discrete Lagrangian by substituting the coarse-grain approximation, given by Eq.~\eqref{eq:coarse_grain_approx}, into Eq.~\eqref{eq:HL_L}, namely
\begin{equation}
\mathbb{L} = \tfrac{1}{2}\mathbb{M}_{11}\dot X_1^2
+ \mathbb{M}_{12}(z)\dot X_1 \dot X_2
+ \tfrac{1}{2}\mathbb{M}_{22}\dot X_2^2
- U(z)
\label{eq:coarse-grain-Lagrangian}
\end{equation}

Thus, we can use the Euler--Lagrange equations for the extremization of the discrete Lagrangian to find the governing system of equations for the centers of the solitons (i.e., the quasi-particles), namely
\begin{subequations}
\begin{align}
\!\!\frac{\dr}{\dr t}\left(\mathbb{M}_{11}\dot X_1\right) + 
\frac{\dr}{\dr t}\left[\mathbb{M}_{12}(z)\dot X_2\right]
+ \mathbb{M}_{12}'(z)\dot X_1\dot X_2
&= \phantom{-} U'(z), \\
\!\!\frac{\dr}{\dr t}\left[\mathbb{M}_{12}(z)\dot X_1\right] +
\frac{\dr}{\dr t}\left(\mathbb{M}_{22}\dot X_2\right) 
- \mathbb{M}_{12}'(z)\dot X_1\dot X_2
&= - U'(z).
\end{align}
\end{subequations}
Finally, recalling that $\mathbb{M}_{ii}$ are constants, $\dot X_i^2 \ll 1$ for $i\in\{1,2\}$, and using the chain rule for differentiation, we can recast the above system into
\begin{subequations}
\label{eq:Mach}
\begin{align}
\mathbb{M}_{11} \ddot X_1 + \mathbb{M}_{12}(z) \ddot X_2 &= \phantom{-} U'(z),\\
\mathbb{M}_{12} (z) \ddot X_1 + \mathbb{M}_{22} \ddot X_2 &= - U'(z).
\end{align}
\end{subequations}

The most conspicuous trait of the above equation of motion is that it is not Newtonian \emph{per se}. It reflects the fact that the inertia of each particle is affected by the presence of another (accelerating) particle in its vicinity\footnote{This should not alarm the reader, as it is now generally accepted that Newton's concept of inertia and his second law of motion are merely the starting point for modern theories (see, e.g., Ref.~\cite{MiltonWillis} and those therein), not the end-all.}. Therefore, it is more appropriate to refer to such pseudo-Newtonian dynamics as \emph{Machean}. Here, by ``Machean dynamics'' we mean those that satisfy a \emph{local} version of \emph{Mach's principle}, which is the hypothesis that the inertia of a body depends not only upon its mass but also upon the quantity, distribution and relative motion of matter in its viscinity \cite{Kleppner}. In Eqs.~\eqref{eq:Mach} this effect manifests itself through the terms whose coefficient is $\mathbb{M}_{12}(z)$. In fact, the \emph{crossmass} $\mathbb{M}_{12}(z)$ gives rise to an interaction between the quasi-particles similar to the interaction due to the potential $U(z)$, so the discrete kinetic and potential energies (i.e., those of the quasi-particles) are not in one-to-one correspondence with their continuous counterparts (i.e., those of the solitons).

Now, it is convenient to render Eqs.~\eqref{eq:Mach} into ``Newtonian form'' by resolving them with respect to the second time derivatives of each quasi-particle, namely
\begin{subequations}
\label{eq:Newton2}
\begin{align}
\ddot X_1 &= \phantom{-} U'(z)\Xi(z)(\mathbb{M}_{22} + \mathbb{M}_{12}) \stackrel{\rm def}{=} \phantom{-} G(z), \\
\ddot X_2 &= - U'(z)\Xi(z)(\mathbb{M}_{11} + \mathbb{M}_{12}) \stackrel{\rm def}{=} - G(z),
\end{align}
\end{subequations}
where
\begin{equation}
\Xi(z) \stackrel{\rm def}{=} \left(\mathbb{M}_{11}\mathbb{M}_{22}- \mathbb{M}_{12}^2\right)^{-1} = \left(64 -64z^2\csch^2 z\right)^{-1}
\end{equation}
for the case when $\mathbb{M}_{11} = \mathbb{M}_{22}$. For this same situation, the ``mathematical'' (or effective) force per unit mass takes the form
\begin{multline}
G(z) = \pm [z - 2\sinh z - (\sinh z - 2z)\cosh z]\\
\times 2\csch^4\bigl(\tfrac{1}{2}z\bigr)\sech^2\bigl(\tfrac{1}{2}z\bigr)/(8\mp8z\csch z),
\label{eq:ode_rhs}
\end{multline}
where the ``$+$'' and ``$-$'' sign still refer to the soliton-soliton and soliton-antisoliton case, respectively. Here, it should be noted that the asymptotic behavior of the right-hand side, as $z\rightarrow 0$, is $G(z)\sim z^{-1}$ for the S-S case and $G(z)\sim z$ for the S-A case. The singularity (i.e., the infinite repulsion) in the S-S case is precisely what prevents the trajectories of the quasi-particles (solitons) from crossing each other.

Finally, we note that Ferguson \& Willis \cite{Willis98} previously performed a number of calculations, in the context of Josephson junctions, similar to the ones presented above. Nonetheless, our results are novel because we propose a very different understanding of the results obtained by the collective-variable method, one that has fundamental physical meaning [embodied by Eqs.~\eqref{eq:Mach}].

\section{Numerical results}
\subsection{Quasi-particle dynamics}
In this section, we provide numerical results, based on the coarse-grain description of the soliton-soliton and soliton-antisoliton interactions, and benchmark them against the \emph{exact} results in \cite{BowtellStuart}. Though a quantitative comparison cannot be made because no scales are provided on the plots in \cite{BowtellStuart}, we can still perform a qualitative comparison.

It is clear that the system given by Eqs.~\eqref{eq:Newton2} can be reduced to a single differential equation governing the evolution of the distance between the quasi-particles, i.e., $X_2 - X_1$. Then, upon obtaining the solution for $X_2 - X_1$, one can go back and solve Eqs.~\eqref{eq:Newton2} for each individual trajectory. For the sake of simplicity, and convenience in comparing the results to the analytical two-soliton solution, we consider here the case of two quasi-particles whose initial phase speeds and positions are equal in magnitude but differ in sign. Consequently, we have
\begin{subequations}
\label{eq:Newton2_IC}
\begin{align}
X_1(t) &= Y(t), &X_2(t) &= -Y(t),\\
X_1(0) &= Y(0) = -x_0, &X_2(0) &= -Y(0) = \phantom{-}x_0,\\
\dot X_1 (0) &= \dot Y(0) = \phantom{-}v_0, &\dot X_2(0) &= -\dot Y(0) = -v_0,
\end{align}
\end{subequations}
and, clearly, $X_2(t) -X_1(t) \equiv -2Y(t)$. Now, we can combine Eqs.~\eqref{eq:Newton2} and \eqref{eq:Newton2_IC} to obtain the following initial-value the problem for $Y(t)$:
\begin{equation}
\ddot Y = G(-2Y), \quad Y(0) = -x_0, \quad \dot Y(0) = v_0. \label{eq:single_pseudoNewton}
\end{equation}
The analytic expression for the effective force (per unit mass) $G(z)$ is known and given by Eq.~\eqref{eq:ode_rhs}, so we can use any standard numerical ODE integration algorithm, e.g., {\sc Mathematica}'s {\tt NDSolve}, to solve the initial value problem for $Y(t)$ given in Eq.~\eqref{eq:single_pseudoNewton}. Naturally, when the function $Y(t)$ is obtained numerically, we can recover the trajectories of the two quasi-particles.

For the initial phase speed, we select a value that falls within the classical limit, namely $v_0=0.1$. Respectively, we choose an initial position of $x_0=6$, which verifies the assumption that (initially) the interaction between the two solitons is insignificant, i.e., they are well-separated.

\begin{figure}
\centerline{\includegraphics[width=7.2cm]{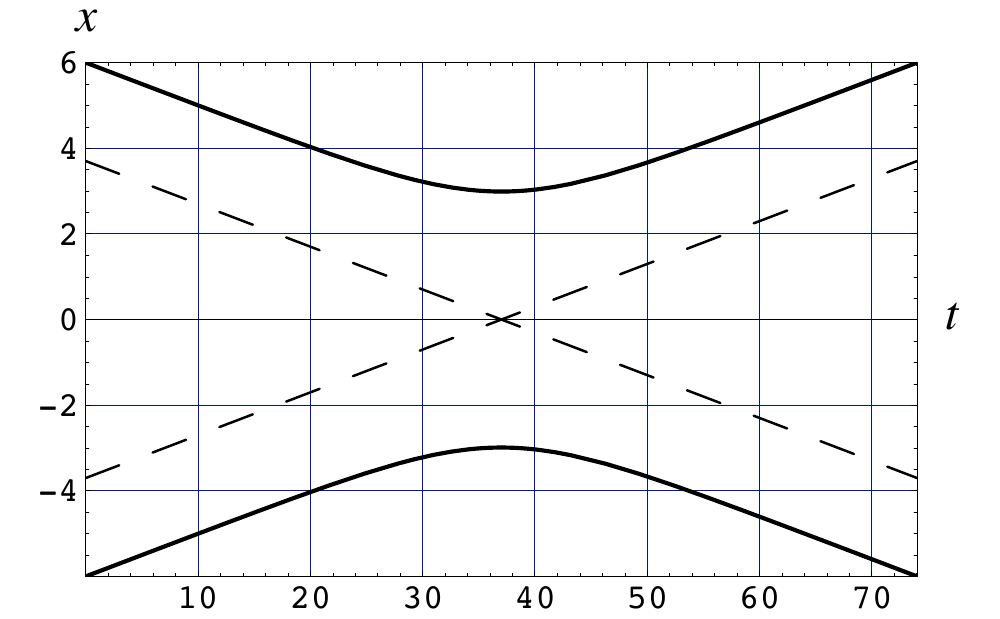}}
\centerline{\includegraphics[width=7.2cm]{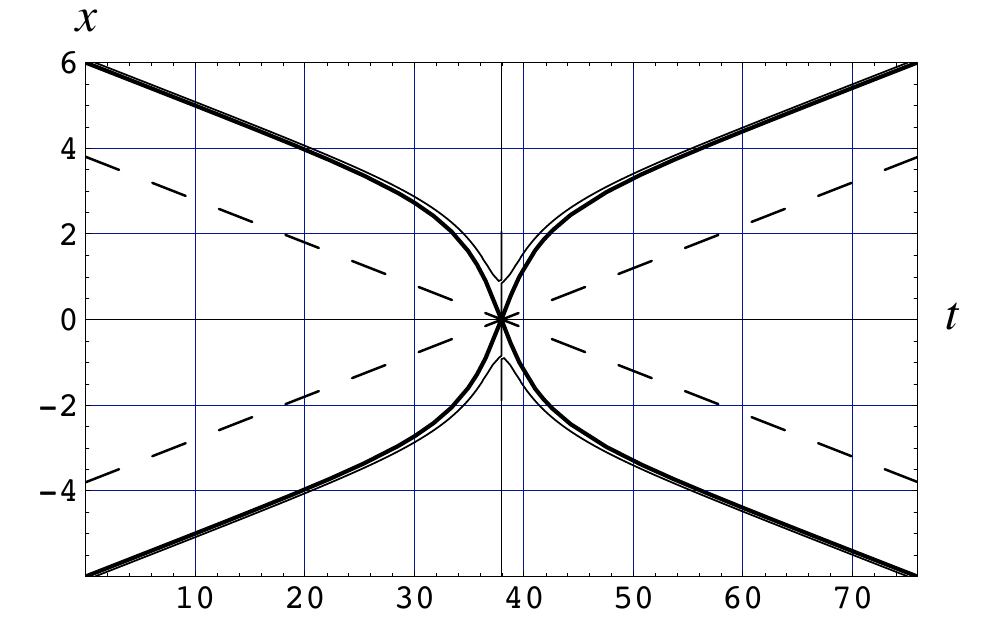}}
\caption{(Color online) The thick solid lines represent the trajectories of the interacting quasi-particles (as calculated from coarse-grain description) are shown for the S-S case (top) and the S-A case (bottom). The dashed lines represent the trajectories of two non-interacting quasi-particles (solitons). And, the thin solid lines represent the location of the inflection points of the  two-soliton solution. \label{fig:traj}}
\end{figure}

Fig.~\ref{fig:traj} presents the results for the S-S (top) and S-A (bottom) solutions. Recall that we defined the centers of the solitons, i.e., the points where $u_x$ has a local maximum (see Fig.~\ref{fig:sol-sol-anti}), as the locations of their corresponding quasi-particles. So, the difference between the straight lines in Fig.~\ref{fig:traj}, which represent the trajectories followed by two non-interacting quasi-particles (solitons), i.e., $X_i=\mp x_0 \pm v_0t$, and the actual trajectories at the time when the quasi-particles (solitons) return to their initial positions gives their phase shifts.

For the S-S case, according to the numerical solution, the time needed for the quasi-particles to return to their starting position [i.e., $t^*\ne 0$ such that $Y(t^*)=Y(0)$] is $t^*=74.0355$. On the other hand, the time needed for the latter quasi-particles to reach the same spatial positions would be $t^*=120$ in the \emph{absence} of interaction between the two quasi-particles. Hence, the interacting quasi-particles return to their initial positions $\Delta t = 120 - 74.0355 = 45.9645$ dimensionless time units earlier than their non-interacting counterparts. Thus, if the non-interacting quasi-particles were to return to their initial positions at the same time as the interacting ones, the former would have to have been separated by a distance $v_0\Delta t = 0.1\times 45.9645 = 4.59645$ smaller than the actual distance of 12 dimensionless length units (see Fig.~\ref{fig:traj}). Therefore, half of this number, $2.29823$, is precisely the phase shift of each quasi-particle. At the same time, substituting the values of the initial phase speed (i.e., $v_0=0.1$) in Eqs.~\eqref{eq:a12}~\&~\eqref{eq:separated}, we can compute the \emph{exact} phase shift of each of the solitons in the full two-soliton solution (for \emph{both} the S-S and S-A case):
\begin{equation}
\delta_1 = -\delta_2 = 2.30259.
\end{equation}
Clearly, the phase shift we obtained from our numerical computation is in very good quantitative agreement with the exact one --- the difference is merely $0.2\%$.

For the S-A case, the solution for the trajectory is shown in Fig.~\ref{fig:traj} (bottom). The time needed for the quasi-particles to swap their positions after their interaction is $t^*=75.9069$, which is $\Delta t = 120 - 75.9069 = 44.0931$ units earlier than their non-interacting counterparts. Therefore, the phase shift of each quasi-particle is $2.20466$, which gives a relative error between the approximate and exact phase shifts of $4\%$. Indeed, in this case, the approximation is supposed to be generally worse than in the S-S case, since in the S-A case the solitons overlap significantly and their velocities get large. This will become clear in the next section.

Thus, the coarse-grain description gives a good, quantitative account of the trajectories of the centers of solitons (i.e., the location of the corresponding quasi-particles). This provides further support for the claim that solitons are indeed particle-like objects and, to a good approximation, they can be considered as point particles obeying pseudo-Newtonian dynamics.

\subsection{Soliton dynamics}
Apart from the significance of the present work as tool for studying the discrete dynamics of quasi-particles corresponding to solitons, our results also show that the coarse-grain description is a practical way of constructing approximate wave-train solutions (two waves in this Letter but, eventually, $N$) of nonlinear wave equations. This is of great importance for non-integrable systems where analytical expressions for wave-train solutions are not available. Here, we outline the quantitative limits of this kind of approximation by comparing our results to the analytical two-soliton solution of the SGE. 

\begin{figure}
\centerline{\includegraphics[width=6.8cm]{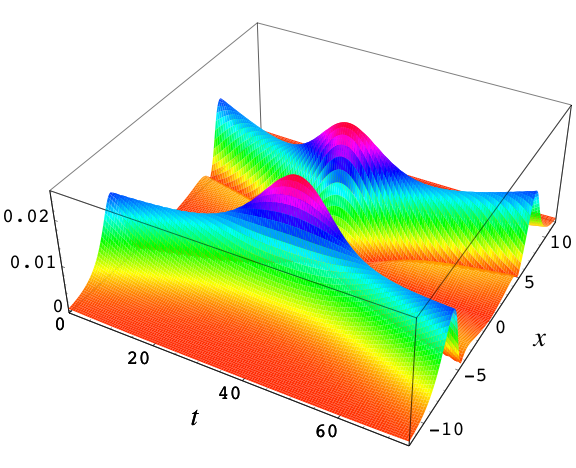}}
\centerline{\includegraphics[width=6.8cm]{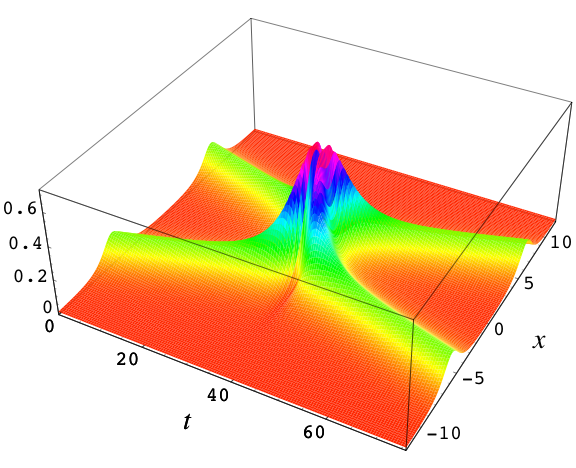}}
\caption{(Color online) Space-time plots of the absolute value of the difference between the exact two-soliton solution and the linear superposition of two single solitons following the calculated trajectories for the S-S case (top) and S-A case (bottom).\label{fig:error}}
\end{figure}

To this end, we compute the difference between the analytical solution, given in Eqs.~\eqref{eq:4pi_kink}, and the coarse-grain approximation, given by Eq.~\eqref{eq:coarse_grain_approx}, presented in the previous section. Fig.~\ref{fig:error} (top) shows the results for the S-S case, where the maximal pointwise error is found to be $\approx 0.028$, which corresponds to a relative error of $0.028/(4\pi)\approx 0.23\%$. In addition, we note that, for the S-S case, the maximum of the phase speed occurs at the initial instant of time, hence the speeds of the solitons (and quasi-particles) remain classical for all times under consideration.

The S-A case is depicted in Fig.~\ref{fig:error} (bottom) and the maximal pointwise error for this case is $\approx 0.73$, which corresponds to a relative error of $0.73/(2\pi)\approx 12\%$. The reason the error for the S-A case is an order of magnitude larger than the errors for the S-S case is that the assumption of classical speeds is no longer justified. In particular, the attraction between the quasi-particles increases their velocities (i.e., the phase speeds of the corresponding solitons) significantly. As a result, we have $\max_t v(t) = \dot X_1(37.9534)=0.675316$, which is no longer small in comparison to unity. Nonetheless, even with the relativistic effects ignored, the error in the approximation (away from the point where the quasi-particles annihilate each other) is small enough so that the coarse-grain description is in good agreement with the exact two-soliton solution.

Here, we note that, for the purposes of these comparisons, we have eliminated any \emph{phase error} by evaluating the two-soliton solution at $t-\vartheta^\pm$, where $\vartheta^\pm$ is the phase error committed by the approximation. Thus, the instant of time when the quasi-particles are closest (S-S case) or pass through (S-A case) each other predicted by the coarse-grain description is the same as the one given by the analytical solution. We found that $\vartheta^+=-0.0660218$ for the S-S case and $\vartheta^-=0.858073$ for the S-A case. Finally, to make sure the quasi-particles and the solitons start at the same place, we had to take, in the analytical solution given in Eqs.~\eqref{eq:4pi_kink}, $x_{1,2}=\mp3.70838$ for the S-S case and $x_{1,2}=\mp3.70954$ for the S-A case.

\section{Conclusions}
An approximate approach to the construction of two-soliton solutions of nonlinear wave equations is discussed. It consist of considering the linear superposition of the non-interacting (or inertially-propagating) shapes of two single solitons following trajectories that are unknown in advance. On the basis of the variational formulation of the field problem, a discrete Lagrangian is derived that governs the trajectories of the corresponding quasi-particles (i.e., the point-particles endowed with certain mass located at the position of the center of a soliton).

The approach was validated using the celebrated sine-Gordon equation, which has analytical two-soliton solutions and the dynamics of the corresponding quasi-particles have been investigate. By numerically integrating the ODEs that govern the trajectories of the quasi-particles, we constructed an approximate two-soliton profile, showing that the quasi-particles obey a pseudo-Newtonian, more precisely Machean, law. Then, the quasi-particle dynamics and the coarse-grain solution were compared to the exact dynamics \cite{BowtellStuart} and the analytical two-soliton solution, respectively. In both cases, excellent quantitative and qualitative agreement was obtained. Furthermore, the present results can easily be extended to the relativistic case by including another collective-coordinate that governs the width of the wave \cite{Rice,Willis98}.

Finally, we emphasize that the collective-co\-or\-di\-nate/var\-i\-a\-ble (or variational approximation) type of approach discussed herein is an effective technique  for studying the two-soliton solutions of equations that are \emph{not} fully-integrable. Therefore, our interpretation of the method --- i.e., the idea of a coarse-grain description --- can shed light on the passage form the continuous to a discrete description of the dynamics (i.e., the quantization) of nonlinear dispersive waves, their relation to the elementary particles and the \emph{concept} of wave-particle duality for more than just the ``classical''  (integrable) equations of soliton theory.

\section*{Acknowledgments}
The authors would like to thank Prof.~C.~R.~Willis for bringing to their attention recent work on the collective-variable approach for two-soliton solutions. All figures appearing in this Letter were generated using the software package {\sc Mathematica} (Version 5.2).

\bibliographystyle{elsart-num}
\bibliography{quasi-particles}

\end{document}